\documentclass[structabstract]{aa}  
\usepackage{graphicx}
\usepackage{txfonts}
\newcommand{\GILDAS}{\texttt{GILDAS}}
\newcommand{\IRAM}{\textrm{IRAM}}
\newcommand{\IRAMthm}{\textrm{IRAM-30m}}
\newcommand{\PdBI}{\textrm{PdBI}}
\newcommand{\CSO}{\textrm{\CSO}}
\newcommand{\ie} {{\em i.e.,}}
\newcommand{\eg} {{\em e.g.,}}
\newcommand{\DCOp}  {\mbox{DCO$^{+}$}}       
\newcommand{\HthCOp}{\mbox{H$^{13}$CO$^{+}$}}
\newcommand{\Jone}{\mbox{$J$=1--0}}
\newcommand{\Jtwo}{\mbox{$J$=2--1}}
\newcommand{\Jthr}{\mbox{$J$=3--2}}

\newcommand{\emm}[1]{\ensuremath{#1}}   
\newcommand{\emr}[1]{\emm{\mathrm{#1}}} 
\newcommand{\unit}[1]{\emm{\, \emr{#1}}}

\newcommand{\kms}   {\unit{km\,s^{-1}}}

\renewcommand{\deg}{\emm{^\circ}}
\newcommand{\Beff}{\emm{B_\emr{eff}}}
\newcommand{\Feff}{\emm{F_\emr{eff}}}

\begin{document}
\title{The ionization fraction gradient across the Horsehead edge:\\
An archetype for molecular clouds\thanks{Based on
observations obtained with the IRAM Plateau de Bure interferometer and
30~m telescope. IRAM is supported by INSU/CNRS (France), MPG (Germany),
and IGN (Spain).}}
                                        
\author{J.R. Goicoechea\inst{1}%
\and J. Pety\inst{2,3}%
\and M. Gerin \inst{3}
\and P. Hily-Blant \inst{4}
\and J. Le Bourlot\inst{5}}%

\offprints{\email{goicoechea@damir.iem.csic.es}}

\institute{Laboratorio de Astrof\'{\i}sica Molecular. Centro de Astrobiolog\'{\i}a. CSIC-INTA.\\
Carretera de Ajalvir, Km 4. Torrej\'on de Ardoz, 28850, Madrid, Spain.
\email{goicoechea@damir.iem.csic.es}%
\and IRAM, 300 rue de la Piscine, 38406 Grenoble cedex, France.\\
\email{pety@iram.fr}
\and LERMA--LRA, UMR 8112, CNRS, Observatoire de Paris and Ecole Normale
Sup\'erieure, 24 Rue Lhomond, 75231 Paris, France.\\
\email{maryvonne.gerin@lra.ens.fr}
\and Laboratoire d'Astrophysique, Observatoire de Grenoble, BP 53, 38041 Grenoble
 Cedex 09, France. \\ \email{pierre.hilyblant@obs.ujf-grenoble.fr}
\and
LUTH, UMR 8102 CNRS, Universite Paris 7 and Observatoire de Paris,  Place J. Janssen
92195 Meudon, France.\\
\email{Jacques.Lebourlot@obspm.fr}
}

\date{Received 10 December 2008 / Accepted 11 February 2009}


  \abstract
  {The ionization fraction (\ie{} the electron abundance) plays a key role
    in the chemistry and dynamics of molecular clouds.}
  {We study the H$^{13}$CO$^+$, DCO$^+$ and HOC$^+$ line emission
   towards the Horsehead, from the shielded core to the UV irradiated
   cloud edge, \ie{} the Photodissociation Region (PDR), as a
   template to investigate the ionization 
   fraction gradient in molecular clouds.}
  {We analyze an IRAM~\textit{Plateau de Bure Interferometer} 
    map of the H$^{13}$CO$^+$ $J$=1--0 line at a $6.8''\times4.7''$ 
    resolution, complemented with IRAM\textit{-30m}  H$^{13}$CO$^+$ and DCO$^+$
    higher--$J$ line  maps and  new HOC$^+$ and CO$^+$ observations.
    We compare self-consistently the observed spatial distribution and line
    intensities with detailed depth-dependent predictions of a PDR model
    coupled with a nonlocal radiative transfer calculation. 
    The chemical network includes deuterated species, $^{13}$C fractionation reactions
    and HCO$^+$/HOC$^+$ isomerization reactions. The role of neutral
    and charged PAHs in the cloud chemistry and ionization balance is
    investigated.}
  {The detection of HOC$^+$ reactive ion towards the
   Horsehead PDR proves the high ionization fraction
   of the outer UV irradiated regions, where we derive a low
   [HCO$^+$]/[HOC$^+$]$\simeq$75--200 abundance ratio.
   In the absence of PAHs, we reproduce the observations 
   with gas-phase metal abundances, [Fe+Mg+...],  lower than
    4$\times$10$^{-9}$ (with respect to H)
   and a cosmic-rays ionization rate of $\zeta$=(5$\pm$3)$\times$10$^{-17}$\,s$^{-1}$.  
   The inclusion of PAHs modifies the ionization fraction gradient and
   increases the required metal abundance.}
   {The ionization fraction in the Horsehead edge follows a steep gradient,
    with a scale length of $\sim$0.05\,pc (or $\sim$25$''$), from [e$^-$]$\simeq$10$^{-4}$ 
    (or $n_e$$\sim$1-5~cm$^{-3}$) in the PDR to a few times $\sim$10$^{-9}$ in the core.
    PAH$^-$ anions play a role in the  charge balance of the cold and neutral gas
    if substantial amounts of free PAHs are present ([PAH]$>$10$^{-8}$).}
  

   \maketitle

\begin{table*}[ht]
\caption{Observation parameters of the PdBI maps shown in Figure~\ref{fig:pdbi-maps}.}
\vspace{-0.2cm}
\begin{center}
{\tiny
\begin{tabular}{ccrclccccccr}
\hline \hline
Molecule  & Transition & Frequency  & Instrument & Config. & Beam   & PA     & Vel. Resol. & Int. Time$^{a}$ & T$_\emr{sys}$ & Noise$^{b,\dagger}$ & \multicolumn{1}{c}{Obs. date} \\
          &            & GHz        &            &         & arcsec & $\deg$ & \kms{}      & hours         & K       & K           & \\
\hline
\HthCOp{} & $ 1 - 0$                            & 86.754288 & PdBI & C \& D & $ 6.8 \times 4.7$ & 13 & 0.2 & 6.5 & 150 & 0.10 & 2006-07 \\
HCO       & $1_{0,1}\,3/2, 2 - 0_{0,0}\,1/2,1$  & 86.670760 & PdBI & C \& D & $ 6.7 \times 4.4$ & 16 & 0.2 & 6.5 & 150 & 0.09 & 2006-07 \\
\hline
\end{tabular}}
\end{center}
\vspace{-0.2cm}
$^{a}$ We observed a 7-field mosaic centered on the IR peak at 
$\alpha_{2000} = 05^h40^m54.27^s$, $\delta_{2000} = -02\deg 28'00''$ (Abergel et al. 2003) with
the following offsets: ($-$5.5$''$, $-$22.0$''$),
(5.5$''$, $-$22.0$''$),
(11.0$''$, 0.0$''$),
(0.0$''$, 0.0$''$),
($-$11.0$''$, 0.0$''$),
($-$5.5$''$, 22.0$''$) and
(5.5$''$, 22.0$''$).
The total field-of-view  is $80.1''\times102.1''$ and the half power primary beam is 58.1$''$.
The mosaic was Nyquist sampled in declination at 3.4~mm and
largely oversampled in right ascension. This maximizes the field of view
along the PDR edge while the oversampling in the perpendicular
direction eases the deconvolution.
On-source time computed as if the source were always observed with 6 antennae.
$^{b}$ The  noise values refer to the mosaic phase center
(mosaic noise is inhomogeneous due to primary beam correction; it 
steeply increases at the mosaic edges).
\label{tab:pdbi-info}
\vspace{0.2cm}

\caption{Observation parameters of the IRAM-30m observations.}
\vspace{-0.5cm}
\begin{center}
{\tiny
\begin{tabular}{lcrlccrclccr}
\hline  \hline
Molecule & Transition    & Frequency  & Instrument & \Feff{}  & \Beff{} & Resol. & Resol. & Int. Time & Noise$^{\dagger}$ & Observing & Obs. date \\
         &               & GHz        &            &          &         & arcsec & \kms{} & hours     &   K               & Mode      & \\
\hline  \hline
HCO$^+$  & \Jone{}       &  89.188523 & 30m/A100   & 0.95     & 0.78    & $27.6''$ & 0.20 & 4.7       & 0.02 & ON-OFF  & 2008 \\
HOC$^+$  & \Jone{}       &  89.487414 & 30m/A100   & 0.95     & 0.78    & $27.5''$ & 0.20 & 4.7       & 0.02 & ON-OFF  & 2008 \\
CO$^+$   & 2, 5/2--1, 3/2& 236.062578  & 30m/A230    & 0.91     & 0.52   & $10.4''$ & 0.20 & 4.7       & 0.05  & ON-OFF  & 2008   \\
\HthCOp{}& \Jone{}       &  86.754288 & 30m/AB100  & 0.95     & 0.78    & $28.4''$ & 0.20 & 2.6       & 0.10 & OTF map & 2006-07 \\
\HthCOp{}& \Jthr{}       & 260.255339 & 30m/HERA   & 0.90     & 0.46    & $13.5''$ & 0.20 & 5.9       & 0.06 & OTF map & 2006 \\
\DCOp{}  & \Jtwo{}       & 144.077289 & 30m/CD150  & 0.93     & 0.69    & $18.0''$ & 0.08 & 5.9       & 0.18 & OTF map & 2006 \\
\DCOp{}  & \Jthr{}       & 216.112582 & 30m/HERA   & 0.90     & 0.52    & $11.4''$ & 0.11 & 1.5       & 0.10 & OTF map & 2006 \\
\hline
\end{tabular}}
\end{center}
\vspace{-0.2cm}
$^{\dagger}$ The noise (in T$_{\emr{mb}}$ scale)  refers to the channel spacing 
obtained by averaging adjacent channels to the velocity resolution given in the tables.
\label{tab:30m-info}
\end{table*}

\section{Introduction}

The electron abundance ([e$^-$]=$n_e/n_\emr{H}$) plays a fundamental role in the chemistry and
dynamics of interstellar gas. The degree of ionization determines the
preponderance of  ion-neutral reactions, \ie{} the main formation
route for most chemical species in molecular clouds
(\cite{her73,opp74}). In addition, the ionization fraction constrains the
coupling of matter and magnetic fields, which drives the dissipation of
turbulence and the transfer of angular momentum, thus having crucial
implications in protostellar collapse and  accretion disks (\eg{}
\cite{bal91}).

High-angular resolution observations of interstellar clouds reveal steep density,
temperature and turbulence gradients as well as sharp chemical variations.
Accordingly, the electron abundance should
vary within a cloud depending on the relative ionizing sources and
prevailing chemistry. 

Rotational line emission of molecular ions such as DCO$^+$ and HCO$^+$ have
been traditionally used to estimate the ionization fraction in molecular clouds 
because $(i)$~they are abundant and easily observable $(ii)$~dissociative
recombination is their main destruction route, and thus their abundances
are roughly inversely proportional to the electron abundance (\eg{}
\cite{gue82,wot82,deB96,wil98}, Caselli et al. 1998, \cite{mar07,hez08}).
On the other hand, the presence of 
reactive ions (species such as HOC$^+$ or CO$^+$ that react rapidly with H$_2$)
is predicted to be a sensitive indicator of high ionization
fraction regions, \eg{} the UV irradiated cloud surfaces (\eg{} \cite{smi02,fue03}).

In order to  constrain the ionization fraction gradient from models, 
the cloud chemistry and physics can not be simplified much because
the charge balance depends on parameters such as the penetration of UV~radiation, 
the cosmic-rays ionization rate ($\zeta$) and the abundance of key species 
(\eg{} metals and PAH).

Compared to other works, in this paper we determine the ionization fraction gradient
by direct comparison of  H$^{13}$CO$^+$ and DCO$^+$  high-angular resolution
 maps and HOC$^+$ pointed observations, with detailed depth-dependent chemical and radiative transfer 
models covering a broad range of cloud physical conditions.
Indeed, the observed field-of-view contains the famous
Horsehead~PDR (the UV illuminated edge of the cloud) and a dense and cold
core discovered by us from  its intense DCO$^+$ line emission
(\cite{pet07}). Due to its simple geometry 
and moderate distance ($d$\,$\simeq$400\,pc), 
the Horsehead PDR and associated core are good templates to
study the steep gradients expected in molecular clouds 
(\eg{} \cite{pet05}, 2007, \cite{goi06,ger09}).

The paper is organized as follows. 
The observations are presented in
Sect.\ref{observations} and the models used to interpret them are
described in Sect.~\ref{models}.  The chemistry of
H$^{13}$CO$^+$, DCO$^+$ and HOC$^+$ (our observational probes of the
ionization fraction) is analyzed in Sect.~\ref{sec-chem-probes}.
In Sect.~\ref{sec-role-of} we investigate
the role of metals, PAHs and  $\zeta$
  on the electron abundance determination.  The main
results and constrains are presented in Sect.~\ref{sec-results} and
discussed in Sect.\ref{sec.discussion}.

\section{Observations}
\label{observations}

\subsection{Observations and data reduction}

Tables \ref{tab:pdbi-info} and \ref{tab:30m-info} summarize the
observation parameters of the data obtained with the \PdBI{} and the \IRAM{}--30m telescope
that we shall study in this work.
 The  \HthCOp\,$J$=1-0 line emission map was first presented in Gerin et al. (2009). 
Frequency-switched, on-the-fly maps (OTF)  obtained at the \IRAMthm{}
were used
to produce the short-spacings needed to complement a 7-field mosaic acquired
with the 6 \PdBI{} antennae in the CD configuration (baseline lengths from
24 to 176~m). Correlator backends were used (VESPA for IRAM-30m  observations).
The high angular resolution PdBI \HthCOp $J$=1-0 map complements our
previous \HthCOp $J$=3-2 and DCO$^+$ $J$=2-1 and 3-2 maps taken with the
IRAM-30m telescope and first presented in Pety et al. (2007).

In this work we present new IRAM--30m deeper integrations 
 in the HOC$^+$, \HthCOp and HCO$^+$ $J$=1-0 lines,
and an upper limit for the CO$^+$ emission towards the PDR
(defined here as the HCO emission peak; \cite{ger09}).
The position switching observing mode was used.
The on-off cycle duration was 1~minute and the off-position offsets were
$(\Delta \alpha, \Delta \delta) = (-100'',0'')$, \ie{}~the {\sc{H\,ii}} region
 ionized by $\sigma$Ori and free of molecular gas emission. 
Position accuracy is estimated to be $\sim$3$''$ for the
30m data and better than $0.5''$ for the PdBI data.
The data processing was done with the \GILDAS\footnote{See
  \texttt{http://www.iram.fr/IRAMFR/GILDAS}} softwares (\eg{}
\cite{pet05b}).  The \IRAMthm{} data were first calibrated to the T$_{A}^*$
scale using the chopper wheel method (\cite{pen73}), and finally converted
to main beam temperatures T$_\emr{mb}$ using the forward and main beam
efficiencies  \Feff{} and \Beff{}  displayed in
Table~\ref{tab:30m-info} (\eg\,\cite{gre98}). 
The  amplitude accuracy for heterodyne observations
with the IRAM--30m telescope is $\sim$10\%.
  PdBI data and
short-spacing data were merged before imaging and deconvolution of the
mosaic, using standard techniques of \GILDAS{}~and used in our previous
works (see \eg{}~\cite{pet05}).

\begin{figure*}[ht]
  \centering %
  \includegraphics[height=0.8\hsize{},angle=270]{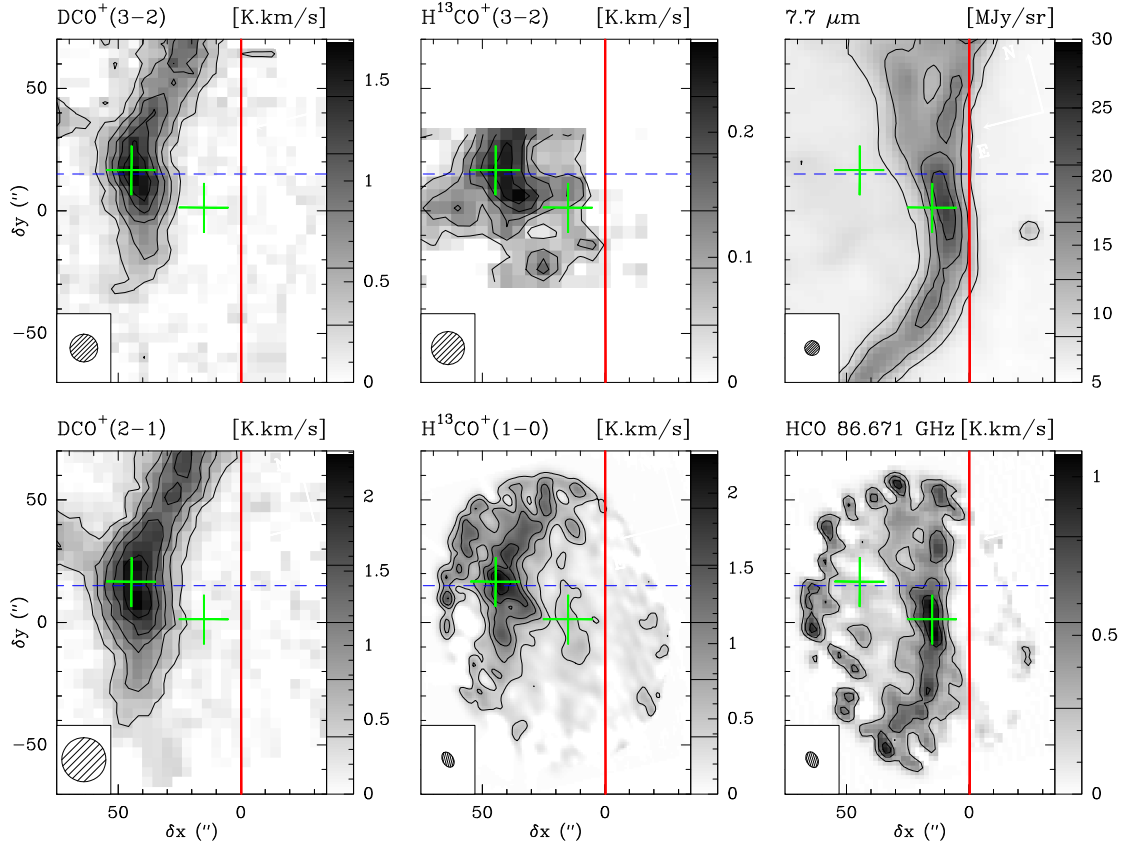} %
   \caption{DCO$^+$ $J$=3--2 and 2--1 (IRAM-30m; from \cite{pet07}), 
      \HthCOp{} $J$=1-0 (PdBI)  and 3-2 (IRAM-30m) line integrated intensity maps, 
      Aromatic Infrared Band emission 
      (ISOCAM, from \cite{abe03}) and HCO (PdBI, from \cite{ger09}).  Maps have been
      rotated by 14\deg{} counter--clockwise around the projection center,
      located at $(\delta x,\delta y)$ = $(20'',0'')$ to bring the
      illuminated star direction in the horizontal direction. The
      horizontal zero has been set at the cloud edge ($\delta$x=0$''$). 
      The \HthCOp{}, DCO$^+$ and HCO
      emission is integrated between 10.1 and 11.1 \kms. Integrated
      intensities are expressed in the T$_{mb}$ scale. 
      Contour levels are displayed on the grey scale lookup tables. The
      red vertical line shows the PDR edge and the green crosses shows two representative
      positions: the \textit{``shielded core''} (the \DCOp{} emission peak
      at $\delta$x$\sim$45$''$; \cite{pet07})
      and the \textit{``PDR''} (the HCO emission peak at $\delta$x$\sim$15$''$; 
      \cite{ger09}). The dashed blue line shows the horizontal cut analyzed in this work.}
  \label{fig:pdbi-maps}
\end{figure*}


\subsection{DCO$^+$ and H$^{13}$CO$^+$ spatial distribution, HOC$^+$ detection}
\label{distribution}

Figure \ref{fig:pdbi-maps} shows H$^{13}$CO$^+$~$J$=1-0, HCO
1$_{0,1}$-0$_{0,0}$ (PdBI) and DCO$^+$ $J$=2-1, 3-2 integrated line
intensity maps (IRAM-30m; \cite{pet07}), as well as the Aromatic Infrared
Band emission (AIB, observed with ISOCAM, \cite{abe03}) that traces the UV
illuminated edge of the cloud, \ie{} the PDR.  The DCO$^+$ emission is
concentrated in a narrow, arclike structure of dense and cold gas behind
the PDR (\cite{pet07}).  Hence, it shows a very different spatial
distribution than the emission of ``PDR tracers'' such as C$_2$H, C$_4$H,
$c$-C$_3$H$_2$ (\cite{pet05}), HCO radicals (\cite{ger09}), 
vibrationally excited H$_{2}$  (\cite{hab05}) or the AIB emission 
(Compi\`egne et al.~2008). The
H$^{13}$CO$^+$ $J$=1-0 emission follows the DCO$^+$ distribution and it
mostly delineates the dense core that coincides with the DCO$^+$ emission
peak.  Nevertheless, while DCO$^+$ is not detected in the illuminated edge,
H$^{13}$CO$^+$ does show a faint emission in the PDR.  Therefore, the small
field-of-view shown in Fig.~\ref{fig:pdbi-maps} contains two different
environments: a warm PDR and a cold core
shielded from the external UV radiation field.  In the following sections
we analyze these emission maps to determine the ionization fraction
gradient in the region.

\begin{figure}[ht]
  \centering %
  \includegraphics[height=0.65\hsize{},angle=-90]{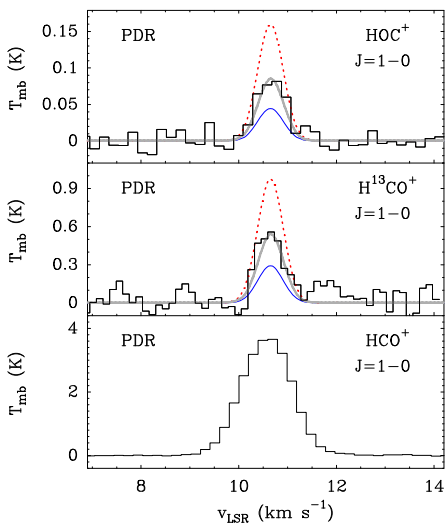} %
   \caption{HOC$^+$ and H$^{13}$CO$^+$ $J$=1--0 lines towards the Horsehead
     PDR (\textit{upper and middle} panels) observed with the IRAM-30m telescope.  
     Solid lines are radiative transfer models with T$_k$=60\,K,
     $n$(H$_2$)=5$\times$10$^4$\,cm$^{-3}$, $n$(H)=500\,cm$^{-3}$ and
     [e$^-$]=5$\times$10$^{-5}$.  Three
     different abundances  are shown,
     grey thick line: [HOC$^+$]=4.0$\times$10$^{-12}$ and 
     [H$^{13}$CO$^+$]=1.5$\times$10$^{-11}$; red dashed line: abundances $\times$2;
     blue thin line: abundances $\div$2.
     For completeness, the HCO$^+$ $J$=1--0 line towards the PDR is also shown  
      (\textit{lower panel}). 
     This transition is very opaque, as shown by the low 
     H$^{12}$CO$^+$/H$^{13}$CO$^+$ $J$=1-0 line intensity ratio ($\sim$7). 
     The resulting line profile is thus broadened
     and it suffers from scattering by low-density
     foreground gas that we do not model here.}
  \label{fig:ions-30m}
\end{figure}

Figure~\ref{fig:ions-30m} shows long integration spectra of the HOC$^+$,
H$^{13}$CO$^+$ and HCO$^+$ $J$=1-0 lines towards the PDR. This is the first
detection of HOC$^+$ reactive ion towards the Horsehead, and adds to
previous detections in interstellar environments with high electron
abundances (\cite{woo83,ziu95,fue03,riz03,sav04,lis04}). H$^{12}$CO$^+$
lines are optically thick, as shown by the low H$^{12}$CO$^+$/H$^{13}$CO$^+$
$J$=1-0 line intensity ratio ($\sim$7), much lower than the expected
$^{12}$C/$^{13}$C$\simeq$60 abundance ratio (\cite{lan90,sav02}) and references
therein). 
The large opacity of H$^{12}$CO$^+$ lines even towards the PDR
justifies the use of H$^{13}$CO$^+$ 
lines as tracers of the HCO$^+$ abundance.

\begin{table}[h]
\caption{Main spectroscopic parameters of the studied lines.}
\vspace{-0.2cm}
\begin{center}
\begin{tabular}{l c c c  c} 
\hline \hline  
Species        & Transition     &  Frequency &  $A_{ij}$            & $E_\emr{upp}$ \\
 &  $J_\emr{upp}-J_\emr{low}$   &    (GHz)   & (s$^{-1}$)           &     (K)      \\ \hline \hline
HCO$^+$        & 1--0           &  89.188523 & 4.2$\times$10$^{-5}$ &  4.3         \\ 
HOC$^+$        & 1--0           &  89.487414 & 2.2$\times$10$^{-5}$ &  4.3         \\ 
\HthCOp{}      & 1--0           &  86.754288 & 3.9$\times$10$^{-5}$ &  4.2         \\
CO$^+$         & 2(5/2)--1(3/2) & 236.062578 & 4.7$\times$10$^{-4}$ &  17.2        \\
\HthCOp{}      & 3--2           & 260.255339 & 1.3$\times$10$^{-3}$ &  25.0        \\
DCO$^+$        & 2--1           & 144.077289 & 2.1$\times$10$^{-4}$ &  10.4        \\ 
DCO$^+$        & 3--2           & 216.112582 & 7.7$\times$10$^{-4}$ &  20.7        \\ 
\hline
\end{tabular}
\label{cs param}
\end{center} 
\end{table} 


\section{Analysis: Models}
\label{models}

In this work we couple the depth-dependent abundances predicted by a PDR model
(for the varying physical conditions prevailing in the Horsehead edge) with detailed
excitation and radiative transfer calculations adapted to the cloud
geometry.  This technique allows us to analyze different chemical models by
direct comparison with observed line intensities. This methodology was
introduced to study our interferometric CS and C$^{18}$O maps of the
Horsehead edge (\cite{goi06}). It enables to observationally
benchmark the abundance gradients predicted by chemical models, even
if it does not produce perfect fits to line profiles in all cloud positions.
 In this paper, we
analyze a horizontal cut of the H$^{13}$CO$^+$ and DCO$^+$ line emission
along the direction of the illuminating star ($\delta y$=15$''$). 
Figure~\ref{fig:pdbi-maps} shows that this
cut (blue dashed line) goes across the DCO$^+$ emission peak ($\delta x$\,$\sim$45$''$),
which we identify as the ``shielded core'', 
and across the HCO emission peak, the ``PDR''
($\delta x$\,$\sim$15$''$).

\subsection{Geometry and density gradient}
\label{subsubsec-dens}

The Horsehead edge has an almost edge-on geometry  with a line-of-sight
depth of $l_{depth}$\,$\simeq$0.1\,pc (\eg~\cite{hab05}) and a spatial
scale  in the plane of the sky of  $\simeq$0.002 pc\,arcsec$^{-1}$.  We determine
the density profile from observations by fitting the 1.2~mm dust continuum
emission (IRAM-30m/MAMBO) along the $\delta y$=15$''$ direction
(\cite{hb05}). In this fit, we adopt a dust opacity per unit (gas+dust)
mass column density of $\kappa_{1.2}$=0.003\,cm$^2$\,g$^{-1}$ at 1.2\,mm
(computed for ``MRN grains'': Mathis, Rumpl \& Nordsieck~1977, see below), 
our best knowledge of the dust
grains temperature (from $\sim$15\,K in the core to $\sim$30\,K in the PDR;
\eg{}~\cite{war06}) and a power-law density profile $n_H(r)$=
$n(\emr{H})$+$2n(\emr{H_2})\propto r^{-p}$, where $r$ is the distance from
the shielded core towards the illuminated
edge of the cloud. Best fits are obtained for a steep density gradient in
the cloud edge ($p$\,$\simeq$3) and a flatter one towards the core
($p$\,$\simeq$0.5).  The turnover point occurs at core radius of
$r$\,$\simeq$0.04\,pc (or $\delta x$\,$\simeq$23$''$ in the maps).
The resulting density gradient used in the photochemical and radiative
transfer models is shown in
Fig.~\ref{fig:depth-mods}.  In the next
sections we constrain the ionization fraction gradient 
in the cloud by comparing synthetic and observed H$^{13}$CO$^+$ and DCO$^+$ 
spectra  along the same cut.

\subsection{Photochemical models}
\label{secPDRcode}

We have updated the \textit{Meudon PDR code} to model our observations 
of the Horsehead.  The
code has been described in detail elsewhere (\eg{}
\cite{jlb93,flp06,goi07}) and benchmarked against other PDR codes
by R\"ollig et al. (2007).
In this section we summarize the most relevant upgrades and
model features for this work.

\subsubsection{UV radiative transfer and dust properties}

The code solves the UV radiative transfer taking into account
dust scattering and gas absorption.
Anisotropic scattering of UV photons by dust grains
is included by explicity calculating the wavelength-dependent
grain albedo and $g$--asymmetry parameters (\cite{goi07}).  This
enables the specific computation of the UV radiation field
(continuum+lines) and thus, the direct integration of consistent
photoionization and photodissociation rates.
We use two types of dust populations: 
$(i)$ a mixture of graphite+silicate
grains and $(ii$) PAHs (see next paragraph). 
More precisely, we adopt a power-law size
distribution ($n(a)$\,$\propto$\,$a^{-3.5}$) with minimum and maximum radius of
$\sim$5 and $\sim$250\,nm respectively (for graphite+silicate grains).
Wavelength-dependent optical properties ($Q$ efficiencies and $g$ factors)
are interpolated from Laor \& Draine (1993) tabulations.  With a standard
gas-to-dust mass ratio ($\sim$100), this grain mixture (``MRN grains'')
reproduces the main characteristics of the standard interstellar extinction curve with
$N_\emr{H}/A_V$\,$=$1.9$\times$10$^{21}$\,cm$^{-2}$ and $R_V$\,$=$3.1.

In order to complete our description of the dust populations, in this work
we have also added smaller aromatic grains.
Observationally, the AIB emission towards the
Horsehead (produced by free PAHs according to the most accepted
theory; \cite{leg84,all85}) clearly separates the {\sc{H\,ii}} region and
PDR (where the emission is bright) from the regions shielded from UV
radiation, where no AIB emission is detected
(\cite{abe03,hab05,com07,com08}).  
However, the size distribution and PAH abundance in dense regions shielded 
from UV radiation  are  uncertain.  
It may vary from ``negligible'', if PAHs coagulate into larger PAH
aggregates (\cite{bou90,rap06}) to ``high'' 
abundances (though they will not be detected in the mid--IR due to the
lack of UV photons to excite them). We used the following PAH properties:
a size distribution with $\sim$0.4 and $\sim$1.2\,nm radii limits
(\cite{des90}) and optical properties from Li \& Draine (2001). 
This size distribution is compatible with PAHs having a mean radius of
$\sim$0.6\,nm and $N_C$\,$\sim$100 carbon atoms assuming
$N_C$\,$\simeq$500\,$a^3$ (\cite{bak94}). The extinction curve and the
efficiency of the photoelectric heating mechanism depend on the mass
fraction put into PAHs (\cite{bak94}). Depending on the PAH
abundance, they contribute to the total dust mass by $\sim$1\%
for [PAH]=10$^{-7}$ and
$\sim$10\% for [PAH]=10$^{-6}$.

\subsubsection{Chemical network and elemental abundances}

Once the UV field is determined at every cloud position, steady-state chemical 
abundances are computed for a given network of chemical reactions.  The model also
computes the temperature profile by solving the thermal balance between the
most important gas heating and cooling mechanisms (\cite{flp06}).  
Our chemical network contains $\sim$160 species and
$\sim$2000 reactions. It includes deuteration, $^{13}$C fractionation
(\cite{gra82}) and HCO$^+$/HOC$^+$ isomerization reactions.
When available, we used the photodissociation rates given by van Dishoeck
(1988), which are explicitly calculated for the Draine's interstellar
radiation field (ISRF).
The most critical reaction rates for our determination
of the ionization fraction
are listed in Table~\ref{tab:rates}.
Most reactions were checked against  \textit{OSU} (E. Herbst and co-workers) and \textit{UDFA}
(\cite{woo06}) networks. Besides, we benchmarked our
network with more extended ones by comparing the
predicted abundances of simple species such as CO and DCO$^+$.

Following Flower \& Pineau des For\^ets~(2003), we have also included
interactions ($\sim$60 reactions) of gas phase species with very small aromatic grains (neutral
PAH and singly charged PAH$^{\pm}$).  In particular, we  take into
account PAH-gas processes such as neutralization reactions of atomic and
molecular cations on PAH$^-$, PAH electron attachment and photodetachment
of PAH$^-$ and PAH by UV photons.  Such processes can play a significant
role in the ionization balance of dense molecular clouds (\eg{}
\cite{lep88,bak98,flo07}, Wakelam \& Herbst 2008, \cite{wol08}). We
have not included larger grains in the  network in order to
isolate the role of PAHs in the gas--phase chemistry. We thus assume
that recombinations of ions with grains are much less frequent than
recombinations with electrons and PAH$^-$.  This is partially justified by
the fact that, according to their size and mass, the fractional
abundance of ``MRN grains'' is low:
$n_\emr{g}/n_\emr{H}$\,$\approx$10$^{-10}$ and their \textit{effective} cross section
per H nucleus  is $(n_\emr{g}/n_\emr{H})\pi a^2$\,$\approx$10$^{-21}$\,cm$^{-2}$
(the product of the grains abundance and the mean grain cross section). 
Grain growth towards the denser parts of
the cloud will result in even lower grain abundances and smaller \textit{effective}
cross sections if the gas-to-dust mass ratio has to be preserved:
\eg{}~$(n_\emr{g}/n_\emr{H})$\,$\approx$10$^{-13}$ and 
$(n_\emr{g}/n_\emr{H})\pi a^2$\,$\approx$10$^{-22}$\,cm$^{-2}$ 
if the grain radii $a$ increase by $\sim$10. Therefore, the resulting lower
abundance of  charged grains and their smaller \textit{effective} 
cross section for ion-grain recombinations will not alter our estimations of
the ionization fraction much.

\begin{table}[h]
\begin{center}
\caption{Key chemical reaction rates$^{\dagger}$ adopted in this work.}
\vspace{-0.2cm}
      \begin{tabular}{ll} 
        \hline \hline
          Reaction    &   Rate  [cm$^3$\,s$^{-1}$]\\
\hline\hline
 \emr{HCO^+  + e^-    \rightarrow CO + H}               &  $2.4 \times 10^{-7} (300\,\emr{K}/\emr{T})^{0.69}\,^{a}$\\
 \emr{HCO^+  + PAH^-  \rightarrow PAH + CO + H}         &  $1.4 \times 10^{-8} (300\,\emr{K}/\emr{T})^{0.50}\,^{b}$\\
 \emr{M^+    + e^-    \rightarrow M + h\nu}             &  $3.7 \times 10^{-12} (300\,\emr{K}/\emr{T})^{0.65}$\\
 \emr{M^+    + PAH^-  \rightarrow M + PAH}              &  $1.0 \times 10^{-8} (300\,\emr{K}/\emr{T})^{0.50}\,^{b}$\\
 \emr{C^+    + H_2O   \rightarrow HCO^+ + H}            &  $8.9 \times 10^{-10} (300\,\emr{K}/\emr{T})^{0.50}$ \\
 \emr{C^+    + H_2O   \rightarrow HOC^+ + H}            &  $1.8 \times 10^{-9} (300\,\emr{K}/\emr{T})^{0.50}$ \\
 \emr{CO^+   + H_2    \rightarrow HCO^+ + H}            &  $7.5 \times 10^{-10}$                               \\
 \emr{CO^+   + H_2    \rightarrow HOC^+ + H}            &  $7.5 \times 10^{-10}$ \\
 \emr{HOC^+  + H_2    \rightarrow HCO^+ + H_2}          &  $3.8 \times 10^{-10}\,^{c}$   \\
\hline
\end{tabular}
\label{tab:rates}
\end{center}
\vspace{-0.3cm}
$^{\dagger}$Rates are from the \textit{UDFA} (\textit{UMIST Database for Astrochemistry}) 
and \textit{OSU} (\textit{Ohio State University}) data bases unless indicated.\\
$^{a}$ Cited in the text as $\alpha$(HCO$^+$), 
it also applies to DCO$^+$ and H$^{13}$CO$^+$.\\
$^{b}$ Rate is from \cite{flo03}.\\
$^{c}$ Rate is from \cite{smi02}.\\
\end{table}
The adopted elemental abundances are shown in
Table~\ref{tab-pdr_std_par}. Low ionization potential heavy \textit{metals}
($\lesssim$8~eV; Fe$^+$, Mg$^+$ or Na$^+$) are all represented by a single
element, ``M$^+$''. In our model, such metals slowly 
recombine with electrons (through radiative recombinations), can be 
photoionized and can exchange
charge with ions and neutrals (including PAHs).  However, they are assumed
to be chemically inert and thus do not form \textit{metallic} molecules
(\eg~\cite{opp74}).  

Once the physical and
geometrical parameters of the cloud are constrained, the only free
parameters in the model are the cosmic-rays ionization rate and the
metal and PAH abundances.
\begin{table}[h]
\caption{Standard conditions and gas-phase elemental abundances.
Abundances, noted [x], refer to H.}
\vspace{-0.2cm}
\begin{center}
\begin{tabular}{l c } 
\hline \hline  
Parameter                                   & Value                                        \\ \hline \hline
Radiation field $\chi$                      & 60 (Draine units)                      \\
Density $n_H(r)$=$n(\emr{H})+2n(\emr{H_2})$ & $\propto r^{-3}$, up to $\sim$\,$2\times10^5$\,cm$^{-3}$   \\
Line of sight depth $l_{depth}$             &  0.1 pc           \\
\emr{[He]}=$n(\emr{He})/n_\emr{H}$          &  $1.00 \times 10^{-1}$                \\
\emr{[O]}                                   &  $3.02 \times 10^{-4}$  \\
\emr{[^{12}C]}                                   &  $1.38 \times 10^{-4}$  \\
\emr{[N]}                                   &  $7.95 \times 10^{-5}$ \\
\emr{[D]}                                   &  $1.60 \times 10^{-5}$                        \\
\emr{[S]}                                   &  $3.50 \times 10^{-6}$ \\
\emr{[^{13}C]}=\emr{[^{12}C]}/60                 &  $2.30 \times 10^{-6}$ \\
\emr{[PAH]}                                 &  variable: 0-$10^{-7}$ \\
\emr{[metals] \equiv [M] \equiv [Fe+Mg+...]}                 &  variable: $10^{-11}$-$10^{-5}$ \\
Cosmic ray ionization rate $\zeta$          &  variable: $10^{-18}$-$10^{-15}$s$^{-1}$ \\ 
\hline
\end{tabular}
\label{tab-pdr_std_par}
\end{center} 
\end{table} 

\subsection{From abundances to spectra: mm radiative transfer}

We use the PDR model predictions (molecular abundance, $n(H_2)$, $n(H)$,
gas temperature and ionization fraction gradients) as input for a nonlocal radiative
transfer calculation able to compute DCO$^+$ and H$^{13}$CO$^+$ line
intensities as a function of cloud position.  Our radiative transfer code
handles edge-on plane-parallel geometry, and accounts for line trapping,
collisional excitation\footnote{PDR-like environments  require to consider
  inelastic collisions with H$_2$, H, He and e$^-$. H$^{13}$CO$^+$, DCO$^+$
  and HOC$^+$ collisional rates with H$_2$, H and He have been scaled from
  those of Flower (1999), while collisional rates with e$^-$ were kindly
  provided by A. Faure and J. Tennyson (see \eg{}~\cite{fau01}).}, and radiative excitation by absorption of
continuum photons. After the level populations are determined in each
modeled slab, emergent line intensities along each line of sight are
computed and convolved with the telescope angular resolution at each
frequency.  A more detailed description is given in Goicoechea et
al.~(2006; Appendix).  Since typical densities in the Horsehead 
($\sim$10$^4$-10$^5$\,cm$^{-3}$) are below the critical densities of the
observed high-dipole moment molecular ions (a few
$\sim$10$^5$-10$^6$\,cm$^{-3}$ for the studied transitions) our approach
allows us to properly take into account non-LTE excitation effects (\eg{}
subthermal excitation), as well as opacity and line profile formation.

\section{Chemistry of the ionization fraction probes}
\label{sec-chem-probes}

\subsection{H$^{13}$CO$^+$ and DCO$^+$ chemistry in the UV shielded core}
\label{subsec-chem-probes-core}

The detection of very bright DCO$^+$ emission towards 
the shielded core (\cite{pet07}) implies cold gas temperatures (T$_k$\,$\simeq$10-20~K)
and thus efficient HCO$^+$ deuterium fractionation (\ie{}
[DCO$^+$]/[HCO$^+$]\,$\gg$D/H).  From our observations we infer 
[DCO$^+$]\,$\simeq8.0\times 10^{-11}$, [H$^{13}$CO$^+$]\,$\simeq6.5 \times 10^{-11}$ and 
thus a [DCO$^+$]/[HCO$^+$]$\simeq$0.02 abundance ratio towards the core peak.
Such gas-phase DCO$^+$ enhancement is achieved
via reaction:
\begin{equation}
\emr{H_3^+ +  HD \rightleftarrows H_2D^+ +  H_2 + \Delta E}
\label{reac:deut1}
\end{equation}
which is endothermic by $\sim$232\,K in the right-to-left direction
(\cite{ger02}), followed by
\begin{equation}
\emr{CO + H_2D^+ \rightarrow DCO^+ + H_2}
\label{reac:deut2}
\end{equation}
which dominates the DCO$^+$ formation in the cold and dense gas.  The
absence of significant DCO$^+$ line emission 
in the PDR is consistent with the
higher temperatures ($>$60\,K) expected in the illuminated edge of the
cloud.  

Besides, the detection of intense H$^{13}$CO$^+$ emission towards the shielded core
and its vicinity (see Fig.~\ref{fig:pdbi-maps})  implies low
ionization fractions. In terms of excitation and opacity effects,
H$^{13}$CO$^+$ is a much more reliable tracer of HCO$^+$ column density
than H$^{12}$CO$^+$ itself (as the latter suffers from very large opacities and
line photon scattering by low-density halos; \eg{}~\cite{cer87}).  In terms
of its chemistry, two main processes dominate the formation of
H$^{13}$CO$^+$ in the low temperature shielded gas:
\begin{equation}
\emr{^{13}CO + H_3^+ \rightarrow H^{13}CO^+ + H_2}
\label{reac:13co-1}
\end{equation}
and isotopic fractionation through
\begin{equation}
\emr{^{13}CO + H^{12}CO^+ \rightleftarrows H^{13}CO^+ +\, ^{12}CO + \Delta E}
\label{reac:13co-2}
\end{equation}
which is endothermic by only $\sim$9\,K in the right-to-left direction
(\cite{lan84}) and competes with dissociative recombination in the destruction of H$^{13}$CO$^+$
where the abundance of electrons is low.  For the physical conditions
prevailing in the shielded core, we predict
[H$^{12}$CO$^+$]/[H$^{13}$CO$^+$] abundance ratios down to $\sim$0.7 times
lower than the elemental [$^{12}$C]/[$^{13}$C] isotopic ratio.  Since both
H$^{13}$CO$^+$ and DCO$^+$ are mainly destroyed by fast dissociative recombination with electrons:
\begin{equation}
\emr{H^{13}CO^+ +  e^- \rightarrow {}^{13}CO +\, H}
\label{reac:hcop}
\end{equation}
\begin{equation}
\emr{DCO^+ +  e^- \rightarrow CO +\, D}
\label{reac:dcop}
\end{equation}
their abundances inversely scale with that of electrons.
In this work we have used a \textit{``standard''}
HCO$^+$ dissociative recombination  rate 
(\ie{}~$\alpha$(HCO$^+$)=$2.4\times$10$^{-7}$(300/T)$^{0.69}$\,cm$^3$\,s$^{-1}$)
recommended in most astrochemical data bases.
We note, however, that there is a certain discrepancy among different
theoretical calculations and measurements   of this key chemical rate
(see discussion by Florescu-Mitchell \& Mitchell~2006 and references therein).
In Sect.~\ref{sec.discussion}
 we discuss the influence  
of adopting a  smaller, \textit{``non standard''} $\alpha'$(HCO$^+$) rate on our results.


\subsection{HOC$^+$ and H$^{13}$CO$^+$ chemistry in the PDR}
\label{subsec-hocp-chem}

In order to extract the [HOC$^+$] and [H$^{13}$CO$^+$] abundances towards
the Horsehead PDR, we have modeled the observed lines
(Fig.~\ref{fig:ions-30m}) using our best knowledge of the prevailing
physical conditions: T$_k$=60-120\,K,
$n$(H$_2$)=5$\times$10$^4$\,cm$^{-3}$, $n$(H)=500\,cm$^{-3}$,
[e$^-$]=5$\times$10$^{-5}$ and a 0.1\,pc line-of-sight depth (or
$N_H$\,$\simeq$3.1$\times$10$^{22}$\,cm$^{-2}$) all accurate within a
factor $\sim$2. From the observed lines we derive the following column
densities: $N$(HOC$^+$)=(1.2--2.5)$\times$10$^{11}$\,cm$^{-2}$ and
$N$(H$^{13}$CO$^+$)=(4.7--7.8)$\times$10$^{11}$\,cm$^{-2}$, which
translates into [HOC$^+$]=(0.4--0.8)$\times$10$^{-11}$ and
[H$^{13}$CO$^+$]=(1.5--2.5)$\times$10$^{-11}$.  This computation assumes
that the HOC$^+$ and H$^{13}$CO$^+$ emission fills the IRAM-30m
  beam. However, HOC$^+$ has not been mapped and its emission could well
arise from the same $\sim$12$''$-width filament where the emission of small
hydrocarbons and HCO radical is concentrated (\cite{pet05,ger09}). In this
case, [HOC$^+$] increases by a factor $\sim$3.  Therefore, we conclude that
the [HOC$^+$]/[H$^{13}$CO$^+$] abundance ratio towards the PDR lies in the range
$\simeq$0.3--0.8.  These values are orders of magnitude higher than the
value expected in the UV shielded gas.

Our chemical models (see next section) reproduce the
[HOC$^+$]/[H$^{13}$CO$^+$] abundance ratio towards the PDR 
but the absolute abundances
derived from observations are larger than those predicted by the model.
The discrepancies between observed and modeled abundances for 
HOC$^+$ and H$^{13}$CO$^+$ have likely a common origin. In particular,
the formation of HOC$^+$ in  UV
irradiated gas is driven by reactions involving C$^+$ and species such as
H$_2$O and CO$^+$ (from  C$^+$~+~OH reaction) that efficiently form at
high temperatures, that is:
\begin{equation}
\emr{C^+ + H_2O \rightarrow HCO^+/HOC^+ + H}
\label{reac:hoc2}
\end{equation}
\begin{equation}
\emr{CO^+ + H_2 \rightarrow HCO^+/HOC^+ + H}
\label{reac:hoc1}
\end{equation}
where reaction \ref{reac:hoc2} predominantly produces HOC$^+$ whereas
reaction \ref{reac:hoc1} has similar branching ratios for the HCO$^+$ and
HOC$^+$ formation (\eg{} \cite{sco97,sav04}). The HOC$^+$ destruction  is
governed by the isomerization reaction:
\begin{equation}
\emr{HOC^+ + H_2 \rightarrow HCO^+ + H_2}
\label{reac:hoc_des}
\end{equation}
Laboratory experiments show that the reaction rate is smaller than
previously thought (\cite{smi02}), allowing interstellar HOC$^+$ to 
exist at detectable amounts.

The intensity peak of the CO $J$=2-1 optically thick lines observed with the PdBI 
(T$_{mb}$$\simeq$60\,K$\lesssim$T$_{ex}$; \cite{pet05}), together with the observed
 CO $J$=4-3/2-1 line ratio (\cite{phi06}),
provide a lower limit to the gas temperature in the PDR  (T$_k$\,$\simeq$60--120\,K).
Temperatures in this range are predicted by the PDR model
but  are not enough to overcome the activation energy
barriers of the neutral-neutral reactions 
leading to the formation of abundant H$_2$O, OH and CO$^+$ (\eg{}~\cite{neu95, cer06}).  
Therefore, our models predict  HOC$^+$ and H$^{13}$CO$^+$ 
abundances lower than observed because their precursor
molecules have low abundances, and
reactions \ref{reac:hoc2} and  \ref{reac:hoc1}  are not efficient enough.

\begin{figure}[h]
  \centering %
  \includegraphics[height=0.75\hsize{},angle=-90]{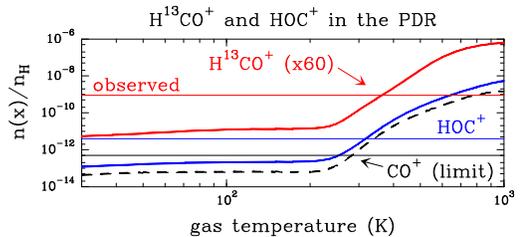} %
   \caption{Predicted H$^{13}$CO$^+$, HOC$^+$  and CO$^+$ 
   peak abundances in the PDR (A$_V$\,$\simeq$0.5--1.5)
   as a function of T$_{gas}$. H$^{13}$CO$^+$ and HOC$^+$ abundances
   (and CO$^+$ abundance upper limit) derived
    from observations towards the PDR are shown with horizontal thin
    lines.} 
  \label{fig:ions-Tgrid}
\end{figure}

We have computed that  gas temperatures around $\sim$350\,K 
are needed to reproduce the observed HOC$^+$ and H$^{13}$CO$^+$ abundances 
in the PDR  through the previous scheme (see  Fig.~\ref{fig:ions-Tgrid}).
Our models of the Horsehead (low UV radiation field) include photoelectric heating
from PAHs and grains but do not predict such a warm gas component even if the PAH
abundance is significantly increased.
However, we do not model the PDR gas dynamics and thus processes such
as shock waves, driven by the expansion of the {\sc{H\,ii}} region
that compress the cloud edge,  may provide
additional gas heating sources to trigger this  \textit{warm chemistry}.
This reasoning is partially consistent with the non detection of CO$^+$
lines, at least at the sensitivity level of our long integration
observation (rms$\sim$50\,mK  in a 0.20\,\kms\, velocity width channel
 or [CO$^+$]$\le$5$\times$10$^{-13}$).

If the gas in the PDR has not gone through such a \textit{warm phase}, reaction \ref{reac:hoc1} has 
to be ruled-out as the main chemical pathway for HOC$^+$ formation and an alternative
formation scenario is required. In this case, we propose that the enhanced HOC$^+$
abundance in the PDR can still be related with the high abundance of C$^+$
(and thus high ionization fraction), but also with grain photodesorption of water-ice
mantles formed in earlier evolutionary stages of the cloud. 



\begin{figure*}[ht]
  \centering %
  \includegraphics[height=1.22\hsize{},angle=0]{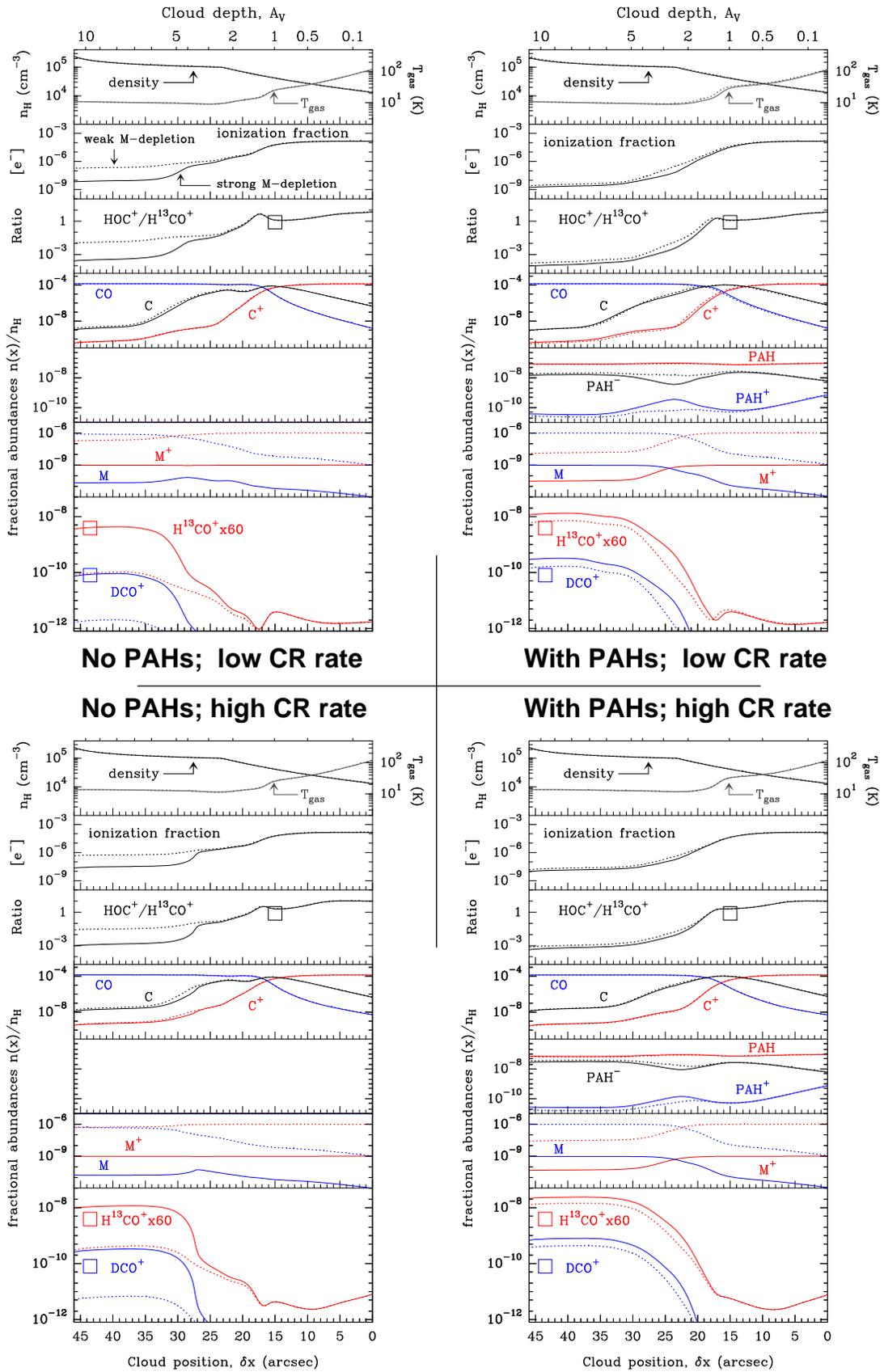} %
   \caption{Grid of chemical models for 2 different metal abundances, 
     \textit{low-metallicity} with \emr{[M]}=10$^{-9}$ (\textit{strong
       metal depletion case}, solid curves) and \textit{high-metallicity}
     with \emr{[M]}=10$^{-6}$ (\textit{weak metal depletion case}, dashed
     curves).  ``Low CR rate'' refers to models with
     $\zeta$=3$\times$10$^{-17}$\,s$^{-1}$ while models with ``high CR
     rate'' refer to $\zeta$=3$\times$10$^{-16}$\,s$^{-1}$.  Models ``with
     PAH'' include PAH-gas interactions in the chemical network (with
     \emr{[PAH]}=10$^{-7}$) while models with ``no PAH'' are pure gas-phase
     models.  The black empty square represents the [HOC$^+$]/[H$^{13}$CO$^+$]
     abundance ratio inferred towards the PDR from observations.  The blue
     and red empty squares represent the DCO$^+$ and H$^{13}$CO$^+$ abundances
     derived towards the shielded core.}
   \label{fig:depth-mods}
\end{figure*}

\clearpage

In this
picture, the low $\chi$/$n$ ratio in the Horsehead ($\sim$10$^{-3}$) will
allow water-ices to be photodesorbed close to the illuminated edge of the
cloud (see predictions by \cite{hol09}), increasing the water vapor
abundance well above the pure--gas phase predictions.  Reaction
\ref{reac:hoc2} will then dominate the HOC$^+$ formation in the PDR.
Taking into account that isomerization, dissociative recombination and
photodissociation contributes to HOC$^+$ destruction, we estimate
that the required water vapor abundance needed to explain the
inferred HOC$^+$ abundance in the PDR is 
[H$_2$O]$\simeq$1800$\times$[HOC$^+$]$\simeq$(0.7-2.2)$\times$10$^{-8}$.
\textit{Herschel Space Observatory} observations  will enable
the detection of C$^+$ and H$_2$O lines in a large sample of PDRs,
confirming whether or not water vapor is abundant at the \textit{edges} of
molecular clouds (\eg{}~\cite{cer06}).\\

\section{Determination of the ionization fraction}
\label{sec-role-of}

Figure~\ref{fig:depth-mods} presents depth--dependent predictions of several
photochemical models across the Horsehead edge. Each model
shows the main physical parameters (density and temperature), the
ionization fraction gradient, the  \DCOp{}, \HthCOp{} and
HOC$^+$ abundances (our observational probes of the ionization fraction)
and the abundances of key chemical species for  
the charge balance in the cloud: CO/C/C$^+$, M/M$^+$, PAH$^-$/PAH/PAH$^+$.

Four sets of models are displayed.  
\textit{Top/bottom} models use a
\textit{low}  ($\zeta$=3$\times$10$^{-17}$\,s$^{-1}$) and
\textit{high} ($\zeta$=3$\times$10$^{-16}$\,s$^{-1}$) cosmic-rays ionization rate
respectively. 
\textit{Left/right} models  exclude and include the effects of PAHs respectively.  
In the latter case,  we include PAHs in the UV radiative transfer  
(as a source of absorption and scattering of UV photons), in the photoelectric heating
 and in the chemical network.
We start the chemistry computation by including neutral PAH alone with
an initial abundance of [PAH]=10$^{-7}$. In each set of models (each panel), 
the only parameter that varies is the abundance of metals:
\textit{high} metallicity with \emr{[M]}=10$^{-6}$ 
(dashed curves) and \textit{low} metallicity with \emr{[M]}=10$^{-9}$ (solid curves). 
The low metallicity case implies a large metal depletion from the gas phase.

In terms of the chemical species observed in this work, 
a salient feature of all models is the constancy of the DCO$^+$/H$^{13}$CO$^+$ abundance ratio 
once the gas is shielded from UV radiation (A$_V$\,$\gtrsim$6). This
feature agrees with the almost
identical spatial distribution of DCO$^+$ and H$^{13}$CO$^+$ emission observed
beyond the PDR (see Fig.~\ref{fig:pdbi-maps}).  This similarity was
already noticed in the lower resolution DCO$^+$ and H$^{13}$CO$^+$
pioneering maps of several molecular clouds (\eg{}~\cite{gue82}).
Besides, the predicted [HOC$^+$]/[H$^{13}$CO$^+$] abundance ratio
towards the PDR is in good agreement with the value inferred
from observations. In this UV irradiated region where the 
 C$^+$ and electron abundances are very high, the HCO$^+$ destruction rate
becomes comparable to the isomerization rate (reaction \ref{reac:hoc_des}).
This implies that the [HOC$^+$]/[H$^{13}$CO$^+$] abundance ratio in the cloud
achieves the highest value in the PDR.

\subsection{The role of ionized carbon and metals}

According to the  ionization fraction gradient all models show two differentiated
environments separated by a transition region: the ``PDR''
(A$_V$\,$\simeq$0-2) and the
``shielded core'' (A$_V$\,$\gtrsim$6). 
The electron density at every cloud position is
given by the difference of cations and anions densities;
\begin{equation} 
  n_e=\sum_i n_i (\emr{cations^+}) - \sum_j n_j (\emr{anions^-}).
\end{equation}
In the PDR, carbon, the most abundant heavy element with a ionization
potential below 13.6\,eV, provides most of the charge:
$n(e^-)$\,$\simeq$\,$n(\emr{C^+)}$. Therefore, the ionization fraction in the PDR is
high, \emr{[e^-]}$\sim$10$^{-4}$ (or $n_e$$\sim$1-5~cm$^{-3}$), and independent of elemental abundances
other than  carbon.

As A$_V$ increases inwards the cloud, the C$^+$ abundance 
decreases by several orders of magnitude and so does the abundance of electrons.
In the shielded core (A$_V$\,$\gtrsim$6), low ionization heavy metal ions (\eg{}~Fe$^+$, Mg$^+$ or Na$^+$) 
determine much of the ionization fraction (\cite{opp74,gue82}).
In the absence of PAHs, abundant molecular ions $m^+$ transfer charge fast to heavy metal 
atoms $M$ through $m^+ + M  \rightarrow m + M^+$ reactions. 
Metal ions recombine orders of magnitude slower than molecular ions 
(Table~\ref{tab:rates}), and thus a large fraction of them is kept ionized
(higher \emr{[M]} implies higher electron abundances). 
Therefore, the ionization fraction in the core is highly dependent on the adopted
metallicity,
and varies from a few $\times 10^{-9}$ for \emr{[M]}=10$^{-9}$, to a few $\times 10^{-7}$ for
\emr{[M]}=10$^{-6}$.

\subsection{The role of PAHs}
\label{sec.rolpahs}

Depending on their abundances, the presence of PAHs can alter
the chemistry and the ionization balance in dense clouds (\eg{}~\cite{lep88}). 
For our adopted abundance of [PAH]=10$^{-7}$ the
right and left panels in Fig.~\ref{fig:depth-mods} shows that the presence 
of PAHs mostly modifies the ionization fraction at A$_V$\,$\gtrsim$2.
If not all PAHs accrete onto bigger grains or coagulate towards cloud
interiors,
\emr{PAH^-} can be abundant through the cloud because the 
radiative electron attachment rate
\begin{equation}
\emr{PAH} + e^- \rightarrow \emr{PAH^-} + h\nu
\label{reac.attach}
\end{equation} 
is high ($\geq$10$^{-7}$\,cm$^3$\,s$^{-1}$), although probably dependent on
the PAH size (\cite{omo86,all89,flo07,wak08}).
In the shielded core PAH$^-$ is destroyed by recombination with 
atomic (M$^+$,...) and molecular cations (HCO$^+$, H$_3$O$^+$,...) 
which are orders of magnitude less abundant than the available cations 
in the PDR (C$^+$, S$^+$,...).
Negative PAH ions thus reach high abundances ([PAH$^-$]$\simeq$2$\times$10$^{-8}$).
  For our choice of PAH parameters, this means that
\textbf{PAH$^-$ can be the most abundant negatively charged species, more
than electrons for A$_V$\,$\geq$5}. 
In addition, recombination
of atomic ions on PAH$^-$ is by far more efficient than the slow radiative
recombination on electrons.  This is a very important point since
\textbf{heavy metal ions are now neutralized at similar rates than
  molecular ions}. As a result, both the abundance of metal cations
and the ionization fraction decreases when PAHs are included, while
molecular ions such as H$^{13}$CO$^+$ and DCO$^+$ increase their 
abundances (see Fig.~\ref{fig:depth-mods} \textit{right panels}).


In the illuminated edge of the cloud, PAH$^-$ is predominantly destroyed by
UV photons through electron detachment,
\begin{equation}
\emr{PAH}^- + h\nu \rightarrow \emr{PAH} + e^-
\end{equation} 
and through recombination with atomic cations which are very abundant in the PDR
(\eg{} \cite{bak98, wol08}). In consequence, the abundance of ions such as
S$^+$ in the PDR decreases with respect to models without PAHs.
This effect is important to determine the elemental abundances and
depletion factors.  Despite the higher PAH$^-$ destruction rates in the
PDR, the high electron density and relatively low UV field in the Horsehead
allows PAH$^-$ to form efficiently through electron attachment
(reaction \ref{reac.attach}). Hence the resulting PAH$^-$ abundance is also high in
the PDR.  On the other hand, the predicted abundance of positively charged
PAH$^+$ in our models is $\sim$500 times smaller than the abundance of
PAH$^-$ (due to 
\begin{figure*}[t]               
  \centering %
  \includegraphics[height=0.35\hsize{},angle=0]{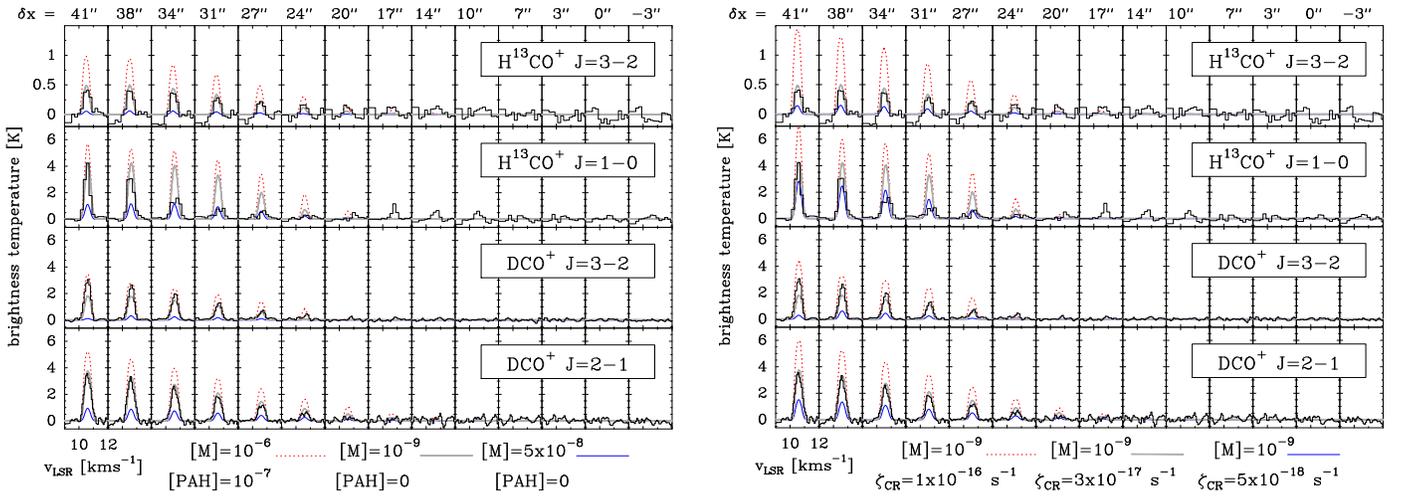}
   \caption{\textbf{\textit{Left}}: H$^{13}$CO$^+$ and DCO$^+$  spectra
     along the direction of the exciting star at $\delta y$=15$''$
     (histograms).  Radiative transfer models using the output of PDR
     models for a fixed cosmic rays ionization rate
     ($\zeta$=3$\times$10$^{-17}$\,s$^{-1}$) and varying metallicities.
     \textit{Thin blue curves} for  \emr{[M]}=5$\times$10$^{-8}$
     and no PAHs; \textit{thick grey curves} for \emr{[M]}=10$^{-9}$ and no
     PAHs; and \textit{dashed red curves} for \emr{[M]}=10$^{-6}$ and
     \emr{[PAH]}=10$^{-7}$. Modeled
     line profiles have been convolved with the appropriate Gaussian beam
     at each observed frequency (the angular resolution for each line
      are quoted in Tables~\ref{tab:pdbi-info} and
      \ref{tab:30m-info}).
     \textbf{\textit{Right}}: Same as previous figure but for a fixed metal 
      abundance (\emr{[M]}=1$\times$10$^{-9}$), no
     PAHs and varying cosmic rays ionization rate $\zeta$.   
     \textit{Thin blue curves} for a model with $\zeta$=5$\times$10$^{-18}$\,s$^{-1}$;
     \textit{thick grey curves} for $\zeta$=3$\times$10$^{-17}$\,s$^{-1}$;
     and \textit{dashed red curves} for $\zeta$=10$^{-16}$\,s$^{-1}$.}
  \label{fig:ions-mods-mtc}
\end{figure*}
fast electronic recombination) and therefore PAH$^+$ don't
seem to play a major role in the ionization balance inside the cloud (see also
\cite{lep88, wak08}).

\subsection{The role of the cosmic-ray ionization rate}

Cosmic rays affect the ionization state and the physics of molecular
clouds, being the dominant source of heating and ionization in the gas
shielded from interstellar radiation fields. Indeed, 
\textit{secondary} UV photons are created in cloud interiors by H$_2$
electronic cascades following H$_2$ excitation by collisions with cosmic
rays (\cite{pra83}). Therefore, cosmic rays maintain a certain ionization
degree in the shielded gas and play a fundamental role in the ion-neutral
chemistry by setting the abundance of key ions (\cite{her73}).

Most studies based on the interpretation of observed molecular ions
set a range of a few 10$^{-17}$ to a few 10$^{-16}$\,s$^{-1}$ for the
cosmic-ray ionization rate (\cite{flp04,vdt06,dal06} and references
therein). However, it is still discussed whether or not $\zeta$ depends on
environmental conditions (\eg{}~galactic center vs. disk clouds) or if it
varies from source to source (\eg{}~from dense molecular cores to more
translucent clouds).  In many ways, PDRs offer an interesting intermediate
medium to analyze the transition between translucent and dark clouds.

In terms of our observations, the DCO$^+$ and H$^{13}$CO$^+$ abundances
directly scale with $\zeta$ in the UV shielded gas. Indeed, 
these  ions are direct products of the
H$_{3}^{+}$  destruction (through reactions \ref{reac:deut2} and \ref{reac:13co-1}), 
and the H$_{3}^{+}$ formation is proportional to $\simeq$\,$\zeta\,n_H$.
However,  $\zeta$ and the metal abundance can not be constrained independently from
the inferred DCO$^+$ and H$^{13}$CO$^+$ abundances 
 since both parameters control the ionization fraction,
and thus the destruction of these ions through  
reactions \ref{reac:hcop} and \ref{reac:dcop}.

\section{Results: observational constraints}
\label{sec-results}

In this section we compare the synthetic and observed H$^{13}$CO$^+$ and
DCO$^+$ spectra as a function of cloud position.  We then explore the range
of metallicities and cosmic-rays ionization rates compatible with the
H$^{13}$CO$^+$ and DCO$^+$ inferred abundances
(see~Table~\ref{tab:abundances}).  The influence of PAHs is also
investigated. We finally compare the [HOC$^+$]/[H$^{13}$CO$^+$] ratio
obtained towards the Horsehead with the values derived in other PDRs.

\subsection{Constraints to the metals abundance}
\label{subsec:metals}

\begin{figure*}[ht]
  \centering %
  \includegraphics[height=0.42\hsize{},angle=0]{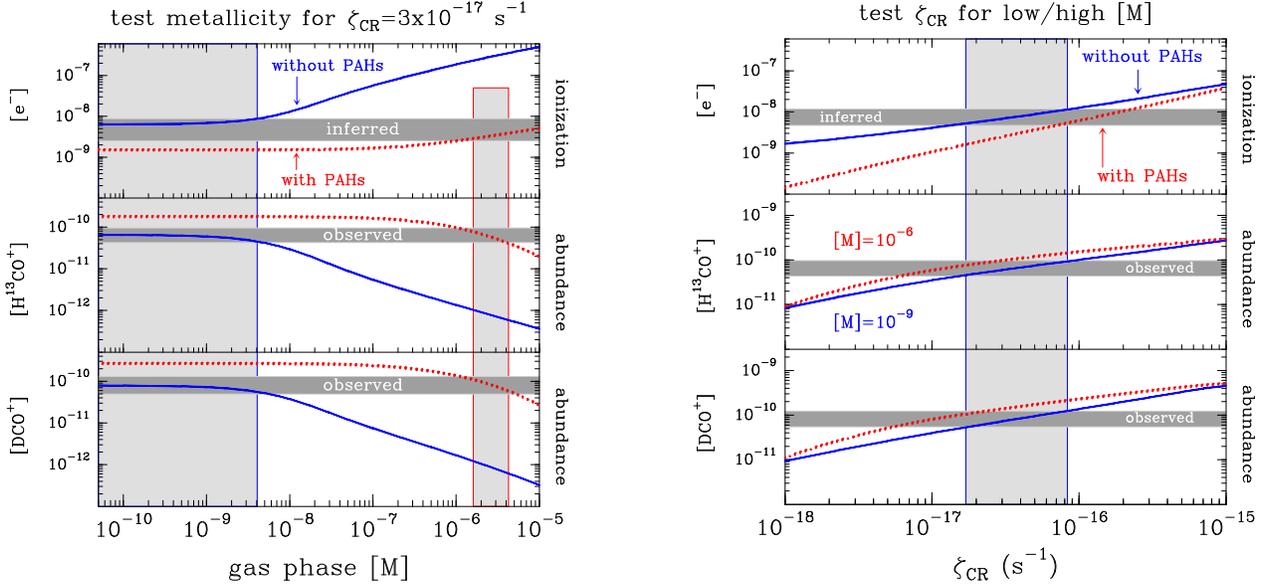} %
   \caption{\textbf{\textit{Left}}: Model predictions for the shielded core  
     (the ``DCO$^+$ peak'' at A$_V$\,$>$10). The ionization rate due to
     cosmic rays is fixed to $\zeta$=3$\times$10$^{-17}$\,s$^{-1}$.  The
     different panels show $(upper)$: the~ionization fraction, $(middle)$:
     the H$^{13}$CO$^+$ abundance and $(lower)$:~the DCO$^+$
     abundance as a function of gas phase metallicity.  Blue-solid curves
     for models without PAHs, and red-dotted curves for models with neutral
     and charged PAHs (\emr{[PAH]}=10$^{-7}$).  Horizontal shaded
     regions show the H$^{13}$CO$^+$ and DCO$^+$ abundances derived from
     observations towards the core, while vertical shaded regions show
     the parameter space compatible with observations.  
     \textbf{\textit{Right}}: Same as previous figure but for a fixed
     low metallicity of \emr{[M]}=10$^{-9}$ (no PAH; blue-solid curves) 
     and a fixed high metallicity of \emr{[M]}=10$^{-6}$ ([PAH]=10$^{-7}$; 
     red-dotted curves).  The different panels show
     $(upper)$: the~ionization fraction, $(middle)$: the H$^{13}$CO$^+$
     abundance and $(lower)$:~the DCO$^+$ abundance as a function of
     the ionization rate due to cosmic rays.}
  \label{fig:Mz-mods}
\end{figure*}

Figure~\ref{fig:ions-mods-mtc}~\textit{left} shows the spectra along the
direction of the exciting star (histograms) and radiative transfer models
using the output of several PDR models for a fixed ionization rate
($\zeta$=3$\times$10$^{-17}$\,s$^{-1}$) and varying metallicities.  In
particular, the model with [M]=10$^{-9}$ (and no PAHs) displays a
notable agreement with both the DCO$^+$ and H$^{13}$CO$^+$ spatial
distribution and with the inferred peak
abundances towards the core (Table~\ref{tab:abundances}).  In addition,
Fig.~\ref{fig:Mz-mods}~\textit{left} shows the predicted ionization
fraction and [H$^{13}$CO$^+$] and [DCO$^+$] abundances at the core peak
(A$_V$\,$>$10) as a function of [M] (blue-solid curves).  These models
(no PAHs, fixed $\zeta$) shows that the \textbf{upper limit metallicity
  compatible with observations is [M]$\leq$ 4$\times$10$^{-9}$, which
  implies a ionization fraction of [e$^-$]\,$=$(7$\pm$1)$\times$10$^{-9}$
  at the core peak}.  Higher metal abundances increase the ionization
fraction (see Fig.~\ref{fig:Mz-mods}~\textit{left}), which translates into
weaker lines than observed (Fig.~\ref{fig:ions-mods-mtc} \textit{left}:
~blue-thin curves).  Therefore, the gas--phase metal abundance is depleted
by $\sim$4 orders of magnitude with respect to the Sun
([M]\,$\simeq$\,8.5$\times$10$^{-5}$; \cite{and89}).  This range of
depletion is similar to that obtained in other prestellar cores such as
Barnard~68 (\cite{mar07}). We shall refer it as the \textit{strong metal
  depletion} case.

\begin{table}[h]
\begin{center}
\caption{Inferred abundances [x]=$N$(x)/$N_\emr{H}$ where
  $N_\emr{H}$=$N$(H)+2$N$(H$_2$).}%
\vspace{-0.2cm}
      \begin{tabular}{lcc} 
        \hline \hline
        Species          &  shielded core          &   PDR \\
                         &  A$_\emr{V}$\,$\geq$6   &   A$_\emr{V}$=0--2\\
                         &  $\delta x$\,$\simeq$45$''$  &   $\delta x$\,$\simeq$15$''$\\
\hline\hline
$N_\emr{H}$ (cm$^{-2}$)  & $5.8 \times 10^{22}$    &  $3.1 \times 10^{22}$\\\hline
\emr{[H^{13}CO^+]}       & $6.5 \times 10^{-11}$   &  $1.5 \times 10^{-11}$\\
\emr{[H^{12}CO^+]}       & $3.9 \times 10^{-9}$    &  $9.0 \times 10^{-10}$\\
\emr{[DCO^+]}            & $8.0 \times 10^{-11}$   &  (--)\\
\emr{[HOC^+]}            & (--)                    &  $0.4 \times 10^{-11}$\,$^{\dagger}$\\
\emr{[CO^+]}            & (--)                    &   $\leq5.0 \times 10^{-13}$\\\hline
\emr{[e^-]}              & $(1-8)\times 10^{-9}$   &  $10^{-6}-10^{-4}$\\
\hline
\end{tabular}
\label{tab:abundances}
\end{center}
$^{\dagger}$ Assuming extended emission. It
  would be 1.2$\times 10^{-11}$ if HOC$^+$ arises from a 12$''$-width
filament as HCO (\cite{ger09}).
\end{table}

The inclusion of PAH interactions implies lower ionization
fractions and enhanced molecular ion abundances (see Fig.~\ref{fig:Mz-mods}~\textit{left}) 
which results in overestimated H$^{13}$CO$^+$ and DCO$^+$ line intensities
towards the core.  Therefore, the abundance of metals (\eg{}~the 
ionization fraction indirectly) has to be increased  to match 
the observed intensities. In particular, Fig.~\ref{fig:ions-mods-mtc}~\textit{left} 
shows that a model with [PAHs]=10$^{-7}$ and [M]=10$^{-6}$ (red-dashed curves)
displays only a factor $<$2 brighter  lines than models with
3 orders of magnitude lower metallicities and no PAHs (grey-thick curves).  
As shown in Fig.~\ref{fig:Mz-mods}~\textit{left},
 \textbf{the inclusion of PAHs makes the range of metal abundances compatible 
with observations much higher now, [M]=(3$\pm$1)$\times$10$^{-6}$}.
The required abundance of metals is
at least a factor $\sim$500 larger than the former case without PAHs. Thus, we
refer it as the \textit{weak metal depletion} case. Note that this metallicity
is still below the gas--phase abundance of Fe+Mg+... elements in the diffuse interstellar
gas (\eg{}~\cite{wol95,how06}) and is consistent with the
incorporation of metals into dust grains in higher density regions (\eg{}~\cite{wol95}). 
On the other hand, the ionization fraction required 
to reproduce the H$^{13}$CO$^+$ and DCO$^+$ abundances
does not change much, [e$^-$]=(4$\pm$1)$\times$10$^{-9}$ at the core peak 
 (red-dashed curves Fig.~\ref{fig:Mz-mods}~\textit{left}). 
Therefore, it is not easy to distinguish between the
\textit{strong metal depletion}  (no PAHs) and \textit{weak metal depletion} (with PAHs) cases
in terms of the ionization fraction. 
The observation of forbidden  lines from metals such as [Fe\,{\sc ii}] towards the
 UV illuminated edges of molecular clouds may help to 
remove this apparent degeneracy.  In one of the few positive cases, the S140 PDR,
the detection of a weak [Fe\,{\sc ii}]26.0\,$\mu$m fine-structure line emission suggests that 
iron is depleted, but with an abundance of $\sim$5$\times$10$^{-8}$ relative to H (\cite{tim96}).
Nevertheless, without mapping and comparing with other PDR tracers, it is 
not obvious to disentangle whether  [Fe\,{\sc ii}] lines arise from the PDR gas
or  from the adjacent (H{\sc ii}) ionization front 
(\eg{}~\cite{mar98,kauf06}).


\subsection{Constraints to the cosmic-ray ionization rate}

Figure~\ref{fig:ions-mods-mtc}~\textit{right} shows again the observed
spectra along the direction of the exciting star (histograms) and radiative
transfer models using the output of PDR models that vary the cosmic
rays rate, without PAHs and a fixed metal abundance of
[M]=10$^{-9}$.  The adopted metal abundance is, within an order of
magnitude, the usual value estimated in prestellar cores 
(\eg{}~\cite{cas99,mar07}) and is compatible with our \textit{strong metal
  depletion case}. 
Therefore, Fig.~\ref{fig:ions-mods-mtc}~\textit{right}  shows the effects
of different ionization rates  directly on the DCO$^+$ and H$^{13}$CO$^+$ line
intensities. For the adopted physical conditions and chemical network, our
observations constrain $\zeta$ within a factor $\sim$2.  In particular,
Fig.~\ref{fig:Mz-mods}~\textit{right} shows the predicted [e$^-$], [H$^{13}$CO$^+$]
and [DCO$^+$] abundances at the core peak (A$_V$\,$>$10) as a
function of $\zeta$, and evidences that in the absence of PAHs,
\textbf{the cosmic rays ionization rate range compatible with the observations of the
Horsehead edge is $\zeta$=(5$\pm$3)$\times$10$^{-17}$\,s$^{-1}$}
(blue-solid curves). If PAHs are included in the chemistry, the metal
abundance has to be increased accordingly to reproduce the observations.
For our [PAH]=10$^{-7}$ model case,
the  required metal abundance needs to be above [M]$\simeq$10$^{-6}$  to
obtain ionization rates above $\zeta$\,$\simeq$10$^{-17}$\,s$^{-1}$
(Fig.~\ref{fig:Mz-mods}~\textit{right}; red-dotted curves).

Note that given the fact that the H$^{13}$CO$^+$  formation in the PDR is not
dominated by the $^{13}$CO + H$_{3}^{+}$ reaction, the H$^{13}$CO$^+$ abundance 
in the UV illuminated gas does not scale with $\zeta$. Therefore, 
we can not further investigate if 
$\zeta$ varies significantly in the transition from diffuse regions to
the shielded core (\eg{}~\cite{mcc03, pad05}).

\subsection{High ionization fraction in the PDR}  
\label{sub-reactive}

The bright [C\,{\sc ii}]158\,$\mu$m (\cite{zho93}) and [C\,{\sc i}]492\,GHz
(\cite{phi06}) fine structure line emission towards the Horsehead PDR,
together with subtle chemical effects such as  the large
[$l$-C$_3$H$_2$]/[$c$-C$_3$H$_2$] linear-to-cyclic abundance ratio
(\cite{tey05}) all show observationally that the ionization fraction is
higher in the UV illuminated edge than towards the cloud interior. 
Nevertheless, all those studies lacked the angular resolution to properly 
measure the ionization fraction gradient.

The abundances of reactive ions such as HOC$^+$ 
are also predicted to be enhanced in the UV illuminated gas, 
where we have shown that
the ionization fraction is high, up to [e$^-$]$\sim$10$^{-4}$, and
that the HOC$^+$  formation is linked to the availability of C$^+$. 
On the other hand, the H$^{13}$CO$^+$ abundance increases as the electron
abundance decreases towards the shielded core.
Therefore, we predict that \textbf{the [HOC$^+$]/[H$^{13}$CO$^+$] abundance
ratio scales with the ionization fraction gradient}, reaching
the highest values in the PDR  (Fig.~\ref{fig:depth-mods}).
In particular, we derive a high [HOC$^+$]/[H$^{13}$CO$^+$]=0.3--0.8
ratio (or a low  [HCO$^+$]/[HOC$^+$]$\simeq$75-200 ratio) towards the PDR,
similar to that observed in other PDRs such as NGC7023  (\cite{fue03}).

\section{Discussion}
\label{sec.discussion}

\subsection{The ionization fraction gradient}

Star forming clouds display different environments depending on the
dominant physical and chemical processes. These environments are, in a
first approximation, similar to those studied here: $(i)$~a low density
\textit{cloud edge}  directly exposed to a 
 UV radiation field from nearby stars; $(ii)$~a \textit{transition} region or
\textit{ridge} where the H$_2$ density increases as the gas temperature
decreases due to the attenuation of the external radiation field. 
UV~photons can still play a significant role depending
on their penetration depths (\eg{}~cloud clumpiness, grain
properties, etc.) and $(iii)$~the denser  \textit{shielded
  cores} that may be externally triggered to form a new generation of stars depending on
their stability against gravitational collapse (\eg{}~\cite{goi08}).

Assuming that the observed field-of-view in the Horsehead nebula is
representative of the above 3 environments, our maps and chemical models
reveal that \textbf{the ionization fraction follows a steep gradient in
  molecular clouds}: from [e$^-$]\,$\simeq$10$^{-4}$ 
at the \textit{edge}
of the cloud (the ``C$^+$ dominated'' region) to a few times
$\sim$10$^{-9}$ in the \textit{shielded cores}. 
The prevailing chemistry and the abundance of
atomic ions such as C$^+$ and S$^+$  determine the ``slope''
of the ionization fraction gradient in the \textit{transition} regions.  
In particular, sulfur (with a
ionization potential of $\sim$10.36\,eV) is a good source of charge
behind the ``C$^+$ dominated'' region.
Advection and time-dependent effects may also
modify the ionization fraction gradient with time. However,  Morata \& Herbst 
(2008) have  shown  models for (uniform) physical conditions  similar to
those in the Horsehead where [HCO$^+$] (and [e$^-$] presumably) does not change
 with time appreciably. 

\begin{figure*}[ht]
  \centering %
  \includegraphics[height=0.34\hsize{},angle=0]{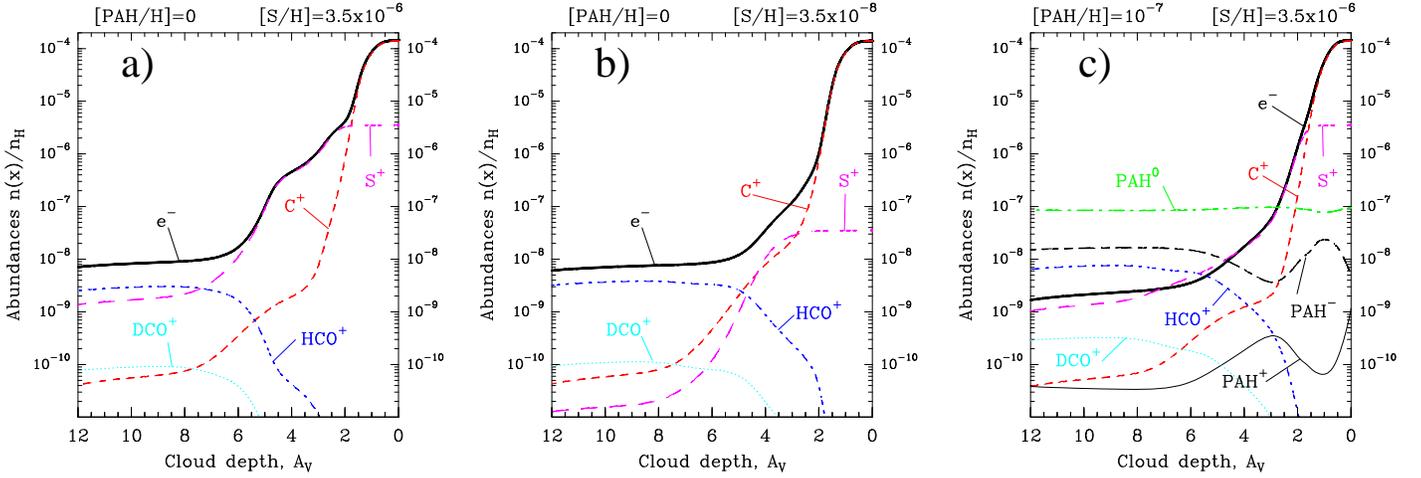} %
   \caption{Derived abundance profiles for the most significant ions studied in this work
     for different PAH and sulfur elemental abundances.  An enhanced UV
     radiation field  60~times the mean ISRF illuminates the cloud from
     the right. The metal abundance ([M]=10$^{-9}$) and the cosmic-ray
     rate ($\zeta$=3$\times$10$^{-17}$\,s$^{-1}$) are fixed in all models.
     $a)$ Model with [PAH]=0 and [S]=3.5$\times$10$^{-6}$ (low sulfur
     depletion).  $b)$ Model with [PAH]=0 and [S]=3.5$\times$10$^{-8}$
     (high sulfur depletion).  $c)$ Model with [PAH]=10$^{-7}$ and
     [S]=3.5$\times$10$^{-6}$.} 
  \label{fig:ions-mods}
\end{figure*}

Figure~\ref{fig:ions-mods} shows the predicted ionization fraction gradient,
with a scale length of $\sim$0.05\,pc (or $\sim$25$''$),
and the main charge carriers for 3 representative models with fixed
\textit{standard} metal abundance (our \textit{strong metal depletion case})
and \textit{standard} cosmic-ray rate.  Panel~\ref{fig:ions-mods}$a$ shows a model without
PAHs and high gas--phase sulfur abundance ([S]=3.5$\times$10$^{-6}$;
\cite{goi06}).  The ionization fraction gradient 
 in the
\textit{core}\,/\,\textit{transition}\,/\,\textit{edge} zones, is mainly
determined by the [HCO$^+$+S$^+$+M$^+$+...]\,/\,[S$^+$]\,/\,[C$^+$]
abundances respectively.  Sulfur ions control the charge balance in the
\textit{transition} layers, and due to their high abundance and slow
radiative recombination rate with electrons, the ionization fraction is
high, a few times 10$^{-7}$, and the gradient is
smooth. This model qualitatively agrees with the observed more extended emission  and narrower 
line-widths of sulfur  recombination lines 
compared to  carbon recombination lines in dark clouds, as well
as with the relatively low ($<$\,10) carbon-to-sulfur recombination lines intensity ratio 
 (\eg{}~\cite{pan78,fal78}). 
All these signatures argue in favor of extended regions with significant amounts of
 gas-phase S$^+$.

Panel~\ref{fig:ions-mods}$b$ shows the same model but sulfur abundances
smaller by two orders of magnitude (strong gas-phase sulfur depletion).  The
ionization fraction in the cloud \textit{core} (A$_V$\,$>$6) and
\textit{edge} (A$_V$\,$<$2) are nearly the same as in the previous high
sulfur abundance model.  However, the lack of abundant S$^+$ in the
\textit{transition} layers decreases the electron abundance considerably,
and makes the ionization fraction gradient much steeper.  
Observational
constraints to the atomic and ionic S abundances from far-IR fine
structure lines or recombination lines, and a careful treatment
of the sulfur chemistry (\ie{} which are the most abundant S-bearing species
as a function of cloud depth?)
are thus required to quantify the S$^+$ abundance at large A$_V$ and
its impact on the charge balance. 

Panel~\ref{fig:ions-mods}$c$ finally shows a model with high sulfur
abundance again but including PAHs (with [PAH]=10$^{-7}$).  As presented
 in Sect.~\ref{sec.rolpahs},
negatively charged PAH$^-$ efficiently form by radiative electron attachment
and their abundance remains high through the cloud. 
 Given the much higher recombination rates of atomic ions
on PAH$^-$ than on electrons, the abundance of atomic ions such as S$^+$ in
the \textit{transition zone}, or M$^+$ in the \textit{shielded cores},
quickly decreases.  Hence, lower ionization fractions (and a much weaker
dependence on the assumed metal elemental abundance) are predicted by the
model with PAHs. This results agrees with theoretical predictions for
UV shielded gas (\cite{lep88,flo07,wak08}). 

In summary, a high abundance of PAHs
\textit{throughout} the molecular cloud (not only in the PDR)
plays a role in the ionization balance and in the abundance of molecular ions,
which affects the determination of elemental abundances (\eg{}~S)
from fractional molecular abundances (\eg{}~HCS$^+$/CS, SO$^+$/SO, etc.).


\subsection{The PAH abundance in UV shielded gas}

The PAH abundance in the  dense and UV shielded gas is far from  being well constrained. 
Different approaches to analyze ISO and
\textit{Spitzer} mid-IR observations towards several PDRs all argue in favor of an
evolution of dust grain sizes: from the illuminated cloud edge
where the emission of PAH bands dominates, to the shielded interiors
where the continuum emission from bigger grains dominates (\cite{rap05},
\cite{ber07}, Compi\`egne et al. 2008).  
It is not trivial to disentangle whether this is a physical effect (\ie{} free PAHs are not
present in the shielded regions) or  an excitation effect (\ie{}~lack
of UV photons). Even if the PAH abundance drastically 
decrease towards cloud interiors, a chemically significant fraction of them may survive.
Unfortunately, while the effects of
grain growth in the UV extinction curve have been modelled by us (\cite{goi07}),
including PAH coagulation/accretion in the chemistry is beyond the scope of this work. 
All we can say at this point is that a better  description of the cloud chemistry may 
be a decreasing 
PAH abundance  gradient or an increasing PAH size distribution towards the 
cloud interior. 
In any case, we have shown than the presence of free PAHs
in molecular clouds modifies the prevailing chemistry.
As a result, the predicted high abundance of PAH$^-$ can dominate the
recombination of metal ions and reduce the ionization fraction.

The presence of abundant free PAHs,  
or large molecules to which electron attach  (\cite{lep88}), can thus be
crucial in determining the coupling of the gas with magnetic fields in
molecular clouds, but also in collapsing cores or in the ``dead'' zones of
protoplanetary disks (magnetically inactive regions where accretion can not occur
if the ionization fraction is very low). According to our models, 
\textbf{the 
PAH abundance threshold required  to affect the metal and electron abundance
determination in the UV shielded gas is [PAH]$>$\,10$^{-8}$}. 
\textit{Herschel} observations might allow the 
identification of specific PAH carriers through their far-IR skeletal  modes 
(\cite{job02}; \cite{mul06}), thus providing clues on their composition and
abundance variations in different environments.

\subsection{``Non standard'' HCO$^+$ dissociative recombination rate}

We conclude by discussing the sensitivity of our determination of the
ionization fraction from  H$^{13}$CO$^+$ and DCO$^+$ abundances.
In particular, we have checked the effects of 
adopting a smaller, \textit{``non standard''}  HCO$^+$ dissociative recombination rate, 
$\alpha '$(HCO$^+$)=$0.7\times$10$^{-7}$(300/T)$^{0.50}$\,cm$^3$\,s$^{-1}$ 
(\cite{she00}, \cite{flo06}). For models without PAHs, 
the predicted H$^{13}$CO$^+$ and
DCO$^+$ abundances increase by a factor $\sim$3 with respect to 
models using the \textit{``standard''} $\alpha $(HCO$^+$) rate 
(Table~\ref{tab:rates}), but
the metallicity required to fit the observed lines has to be
increased  to [M]$\simeq$5$\times$10$^{-8}$ and the
predicted ionization fraction 
increases to [e$^-$]$\simeq$5$\times$10$^{-8}$ in the core. This value
should be regarded as the upper limit of our determination.
On the other hand, the influence of  $\alpha '$(HCO$^+$)
in models with PAHs is less important.  It also requires
high metallicities 
to fit the observed intensities (\textit{weak metal depletion} case), 
but the predicted [e$^-$] in the shielded core remains low (below $\sim$10$^{-8}$).

\section{Summary and conclusions}
\label{summary}

We have presented  the first detection of HOC$^+$ reactive ion towards the Horsehead PDR. 
Combined with our previous IRAM-PdBI H$^{13}$CO$^+$ $J$=1--0 (\cite{ger09})
and IRAM-30m H$^{13}$CO$^+$ and DCO$^+$ higher-$J$ line maps 
(\cite{pet07})
we performed a detailed
analysis of their chemistry, excitation and radiative transfer to constrain
the ionization fraction as a function of cloud position. 
The observed field contains 3 different
environments: 
$(i)$~the UV~illuminated
\textit{cloud edge}, $(ii)$~a \textit{transition} region or \textit{ridge} and $(iii)$~a
dense and cold shielded \textit{core}.  We have presented a study of the
ionization fraction gradient in the above environments, which can be
considered as templates for most molecular clouds.  Our main conclusions
are the following:

\begin{enumerate}
  
\item The ionization fraction follows a steep gradient, 
 with a scale length of $\sim$0.05\,pc ($\sim$25$''$), from
  [e$^-$]\,$\simeq$10$^{-4}$ ($n_e$$\sim$1-5~cm$^{-3}$) at the cloud \textit{edge} (the ``C$^+$
  dominated'' regions) to a few times $\sim$10$^{-9}$ in the
  \textit{shielded core} (with ongoing deuterium fractionation).  Sulfur, metal
  and PAH ions play a key role in the charge balance at different cloud
  depths.

\item The  detection of HOC$^+$ towards the PDR, and 
  the  high  [HOC$^+$]/[H$^{13}$CO$^+$]$\simeq$0.3-0.8 abundance ratio inferred,   
  proves the high ionization fraction in the UV irradiated gas.
  However, the H$^{13}$CO$^+$ and HOC$^+$ abundances derived from observations 
  are larger than the PDR model predictions.
  We propose that either the gas is/was warmer than predicted  or that
  significant water ice-mantle photodesorption is taking place
  and HOC$^+$ is mainly formed by the C$^+$\,+\,H$_2$O reaction.

\item The ionization fraction in the shielded core depends on
  the metal abundance and on the cosmic-rays ionization rate. 
  Assuming a \textit{standard} rate of
  $\zeta$=3$\times$10$^{-17}$\,s$^{-1}$ and pure gas-phase chemistry (no
  PAHs), the metal abundance has to be lower than 4$\times$10$^{-9}$
  (strong metal depletion).  Conversely, assuming a \textit{standard} metal
  abundance of [M]=10$^{-9}$, our observations can only be reproduced
  with $\zeta$=(5$\pm$3)$\times$10$^{-17}$\,s$^{-1}$.

  \item The inclusion of  PAHs  
  modifies the ionization fraction gradient and decreases the metal
  depletion required to reproduce the observations  if
  [PAH]$>$ 10$^{-8}$ (\ie{} if not all PAHs coagulate/accrete onto bigger grains).   
  In such a case, PAH$^-$ acquire large abundances also in the shielded gas.
  Recombination of atomic ions on PAH$^-$ is much more efficient
  than on electrons and thus metal ions are neutralized at similar rates
  than molecular ions.  For [PAH]=10$^{-7}$, the metal abundance consistent
  with observations increases to [M]=(3$\pm$1)$\times$10$^{-6}$
  (still below the heavy metals abundance in the diffuse ISM).
  

\end{enumerate}

\begin{acknowledgements}
We thank the IRAM staff for their support during observations
and D. Talbi and B. Godard for useful advice regarding the HCO$^+$ dissociative recombination
rate.
Inelastic collisional rates of HCO$^+$ with electrons were kindly provided
by A. Faure and J. Tennyson. We also thank M. Walmsely for several interesting
comments.
We acknowledge the use of \textit{OSU} 
(http://www.physics.ohio-state.edu/$\sim$eric/research.html)
and \textit{UDFA} (http://www.udfa.net/) 
chemical reaction data bases.
We finally acknowledge financial support from
CNRS/INSU research programme PCMI.
JRG is supported by a \textit{Ram\'on y Cajal} research contract from the Spanish MICINN
and co-financed by the European Social Fund. 

\end{acknowledgements}

\end{document}